\documentclass[conference]{IEEEtran}
\IEEEoverridecommandlockouts
\usepackage{url}
\usepackage{svg}
\usepackage{cite}
\usepackage{amsmath,amssymb,amsfonts}
\usepackage{algorithmic}
\usepackage{graphicx}
\usepackage{textcomp}
\usepackage{xcolor}
\def\BibTeX{{\rm B\kern-.05em{\sc i\kern-.025em b}\kern-.08em
    T\kern-.1667em\lower.7ex\hbox{E}\kern-.125emX}}
\begin{document}
\bibliographystyle{plain}

\title{An Efficient Multitask Learning Architecture for Affective Vocal Burst Analysis}

\author{

\IEEEauthorblockN{Tobias Hallmen}
\IEEEauthorblockA{\textit{Chair for Human-Centered Artificial Intelligence} \\
\textit{Augsburg University}, Germany \\
tobias.hallmen@uni-a.de}
\and
\IEEEauthorblockN{Silvan Mertes}
\IEEEauthorblockA{\textit{Chair for Human-Centered Artificial Intelligence} \\
\textit{Augsburg University}, Germany \\
silvan.mertes@uni-a.de}
\and
\IEEEauthorblockN{Dominik Schiller}
\IEEEauthorblockA{\textit{Chair for Human-Centered Artificial Intelligence} \\
\textit{Augsburg University}, Germany \\
dominik.schiller@uni-a.de}
\and
\IEEEauthorblockN{Elisabeth André}
\IEEEauthorblockA{\textit{Chair for Human-Centered Artificial Intelligence} \\
\textit{Augsburg University}, Germany \\
andre@uni-a.de}

}

\maketitle

\begin{abstract}
Affective speech analysis is an ongoing topic of research. A relatively new problem in this field is the analysis of vocal bursts, which are nonverbal vocalisations such as laughs or sighs. Current state-of-the-art approaches to address affective vocal burst analysis are mostly based on wav2vec2 or HuBERT features. In this paper, we investigate the use of the wav2vec successor data2vec in combination with a multitask learning pipeline to tackle different analysis problems at once. To assess the performance of our efficient multitask learning architecture, we participate in the 2022 ACII Affective Vocal Burst Challenge, showing that our approach substantially outperforms the baseline established there in three different subtasks.
\end{abstract}


\section{Introduction}


When developing systems to automatically recognise emotions from a given input, current state-of-the-art models often rely on large amounts of manually annotated training data. 
While humans have an innate capability of recognising a universal set of emotions \cite{ekman1992argument}, recent studies show that socio-cultural factors, such as culture or language impact the perceived authenticity and accuracy of a recognised emotion \cite{cosme2022cultural, chronaki2018development}. 
Therefore, it makes sense to take this information into account when developing machine learning models for automatic emotion recognition. 

A particularly interesting problem arises when differences in culture are not apparent through linguistic features, but only through prosodic ones. 
This is the case, for example, when emotions are not expressed through verbal speech, but rather through non-verbal vocal expressions, called \emph{Vocal Bursts}.
Here, cultural differences are not immediately obvious.
However, it is still conceivable that humans develop the expression of vocal bursts differently depending on their social and cultural environment.

The \emph{2022 ACII Affective Vocal Burst Challenge} introduces the problem of understanding emotion in vocal bursts, like grunts, laughs or gasps.
The introduced HUME-VB \cite{Cowen2022HumeVB} large-scale dataset consists of samples of emotional human vocalisations annotated with respect to emotion and culture with vocal bursts originating out of 4 countries.
Thus, it is well suited to explore methods for incorporating socio-cultural aspects into automatic emotion recognition, which is the aim of this paper.
The challenge consists of four subtasks: (i) In the \emph{High} task, the intensity of 10 emotions has to be identified, (ii) in the \emph{Two} task, the two-dimensional valence/arousal representations have to be assessed, (iii) in the \emph{Culture} task, participants have to predict the intensity of 10 emotions while paying respect to culture-specific groundtruths, and (iv) in the \emph{Type} task, the type of expressive vocal bursts (\emph{Cry, Gasp, Groan, Grunt, Laugh, Pant, Scream, Other}) has to be recognised.

To address these tasks, we focus on a combination of feature and multitask learning to develop one coherent model that does not require explicit modelling of features or context knowledge.\footnote{Code available at \url{github.com/hcmlab/acii-2022-vb-challenge}.} 

The continued success of deep learning approaches to audio recognition tasks raises the question of whether hand-crafted features are still necessary or automatic learning of such features is a better way to go \cite{wagner2018deep}. 
Recently, Baevski et al. \cite{baevski2020wav2vec} proposed the wav2vec2 framework that is able to learn meaningful feature representations from the raw waveform of an audio signal that can outperform state of the art approaches for the task of automatic speech recognition. 
Especially for voice emotion recognition, recent work often relies on such embeddings as a backbone for the respective models
\cite{wagner2203dawn, wang2021fine, pepino2021emotion, chen2021exploring}.
Data2vec~\cite{baevski2022data2vec}, a generalised, abstracted derivate of wav2vec2~\cite{baevski2020wav2vec} is a modality-agnostic approach to learning latent representations in a self-supervised way. 
Data2vec, compared to wav2vec2, directly predicts contextualized latent representations without quantization.
For learning audio representations, data2vec is pretrained on the Librispeech ASR corpus~\cite{panayotov2015librispeech}, respectively Libri-light ASR benchmark~\cite{kahn2020libri} for the large version, and released to the public\footnote{\url{github.com/facebookresearch/fairseq/tree/main/examples/data2vec}}. 

Multitask learning for vocal bursts has already been a task in a previous iteration of the challenge \cite{BairdExVo2022}.
There, participants were asked to predict the expression of 10 emotions along with the age and native country of the speaker at the same time. 
Sharma et al. \cite{sharma2022self} approached the task by experimenting with various encoder front ends as well as handcrafted features. 
They found that using the HuBERT model \cite{hsu2021hubert}, which is closely related to the wav2vec architecture and training approach as a backbone yielded the best performance.
Purohit \cite{purohit2022comparing} et al. compared various embeddings that have been either trained using self-supervision or directly in a task-dependent manner. 
They found that overall, the self-supervised embeddings are outperforming the task-dependent ones, which, supports the choice of data2vec for our experiments. 
Anuchitanukul et al. \cite{anuchitanukul2022burst2vec} also rely on wav2vec and HuBERT backbones to extract embeddings for their multitask training system.
They further utilise an adversarial training approach to disentangle the input representations into shared, task-specific ones.
Their experiments showed that the wav2vec-based model performs best, but using ensemble techniques to combine multiple variations of their wav2vec and HuBERT models can achieve even higher performance. 
All three works indicate that self-supervision in general and wav2vec specifically are building a good foundation for the task at hand and confirming our choice of data2vec as the successor of wav2vec. 

\section{Approach}

Since vocal bursts are not ``speech made out of words'', but rather ``speech made out of vocalised emotions'', the not fine-tuned version of data2vec is used.

To follow data2vec's agnostic approach, no or as few assumptions as possible are posed for the downstream supervised fine-tuning. To use all the provided labels, multi-task learning is applied in a self-learning way~\cite{kendall2018multi}, which approximates optimal task weights by learning task uncertainty, including the non-challenge task \emph{Country} -- classifying the vocal burst's origin.

\begin{figure}[ht]
\centering
\includegraphics[height=8.6cm]{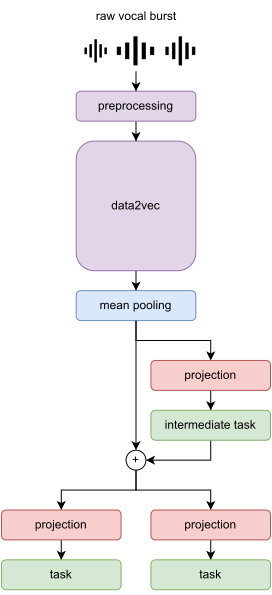}
\caption{Overview of the net architecture.}
\label{fig:net}
\end{figure}

In Figure~\ref{fig:net}, the net architecture is depicted. The raw audio is fed into the pretrained data2vec model, including its preprocessor. Variable sequence lengths are zero-padded to the longest seen sequence. Because of the attention mask, the extracted features vary in length. Therefore, these are mean-pooled and fed into downstream projection layers. To investigate the question of whether knowing certain tasks before predicting other tasks is helpful, the intermediate tasks are separated and their prediction is fed along with the extracted features to the remaining tasks.

The projection layers reduce the output dimension of data2vec (768 base, 1024 large) to 256, apply GELU~\cite{hendrycks2016gaussian}, then further reduce the dimension down to the five task's required dimensions  (high: 10, culture: 40, two: 2, type: 8, country: 4). The tasks' layers then apply softmax for classification and compute the loss via cross-entropy using inversely proportional class weights. For regression, we apply sigmoid and compute the loss using the concordance correlation coefficient. The losses are then linearly combined through the learnt optimal weights.

The whole net has a size of 360MB and is trained in a two-stage matter: First, freeze data2vec and train only the tasks with their projection layers. Second, unfreeze data2vec and fine-tune the whole net.
Both stages are trained for 30 epochs, early stopping with a patience of 2, batch size of 32 for the base version and 24 for the large version, and optimiser AdamW~\cite{loshchilov2017decoupled}. For the first stage, the optimiser uses defaults, for the second one, the learning rate is initialised at $4 \cdot 10^{-5}$ and follows a cosine schedule with a warmup of 1 epoch.

Training is done on a single Nvidia A40 GPU and takes 5 to 6.5 hours for the base version of data2vec, and 5.5 to 8.5 hours for the large version.

\vspace{-0.15cm}




\section{Experiments \& Results}

In our experiments, we varied certain details of the network architecture as listed below. The results for the validation set are shown in Table \ref{tab:results_val}, additionally to the official baseline results presented in \cite{BairdA-VB2022}. Note, that the single experiments were conducted iteratively, carrying the best configuration of the previous experiments over to the next set of experiments.

\emph{Intermediate Tasks.} Following the challenge's motivation that vocal bursts depend on the country of origin and their type to assess the conveyed emotions like a rater of the same origin would we tried three different variants: (i) \emph{Country} and \emph{Type} as intermediate tasks before predicting the remaining three (\emph{2/3}).
(ii) Only \emph{Country} as an intermediate task (\emph{1/4}), assuming the type depends less on the country.
(iii) No intermediate tasks, i.e., predict all tasks simultaneously (\emph{0/5}). (iii) performed best.

\emph{Task Loss.} To investigate, why the task \emph{Two} performs so badly, we experimented with replacing the CCC loss by (i) mean squared error (\emph{MSE}) and (ii) mean absolute error (\emph{MAE}). (iii) We excluded the artificially most certain task \emph{Two} from training (\emph{-Two}). (iii) turned out to be the best option.

\emph{Weighting.} Since the cross-entropy losses in tasks \emph{Country} and \emph{Type} have inverse proportional class weights and the other tasks do not, we experimented with (i) removing these class weights from the training (\emph{-CW}), so every sample's tasks are unweighted, (ii) adding sample weights inversely proportional to \emph{Country} (\emph{+SW}), thereby weighting all tasks, and (iii) keeping the class weights but removing the sigmoid activation from the last layer(\emph{-SM}) and clamping the output to $[0, 1]$ in the regression tasks \emph{High} and \emph{Culture}. We could substantially boost the network's performance by applying (iii).

\emph{Fine-Tuning Splits.} To investigate if fine-tuning the self-supervised learnt audio representations using labelled speech helps, we experimented with different versions of the base network, where each version was fine-tuned after pretraining using connectionist temporal classification (CTC) loss on a different amount of LibriSpeech: (i) 10 minutes (\emph{B10m}), (ii) 100 hours (\emph{B100h}), (iii) 960 hours (\emph{B960h}). Version (iii) performed best.

\emph{Network Size.} At last, we replaced the base network with (i) the \emph{Large} (\emph{L}) version and (ii) the \emph{Large Fine-tuned} (\emph{L960h}) version. Also, we experimented with completely dropping the \emph{Country} task (\mbox{\emph{-Country}}). However, no improvement could be observed in any experiment here.

As can be seen in Table \ref{tab:results_val}, for the \emph{High} and \emph{Culture} tasks, experiment \emph{B960h} worked best on the validation set.
For the tasks \emph{Type} and \emph{Country}, experiment \emph{-SM} performed best on the validation set. To obtain results on the test set, predictions from both models (\emph{B960h} and \emph{-SM}) were submitted to the challenge organisers. The tasks \emph{Country} and \emph{Two} were omitted as non-challenge, respectively unsuited representations for valence/arousal estimation. The results on the test set are shown in Table \ref{tab:results_test}. As can be seen, our results outperform the baseline in all three submitted tasks.



\begin{table}[ht]
\centering
\caption{Results on the validation split of the A-VB challenge dataset.}
\begin{tabular}{lccccc}
\hline
         & High          & Culture       & Two           & Type          & Country       \\
Method   & CCC           & CCC           & CCC           & UAR           & UAR           \\ \hline
Baseline & .564          & .436          & \textbf{.499} & .412          & --            \\ \hline
2/3      & .639          & .625          & .252          & .562          & .631          \\
1/4      & .628          & .614          & .245          & .542          & .603          \\
0/5      & .650          & .636          & .254          & .564          & .633          \\ \hline
MSE      & .624          & .598          & .244          & .553          & .591          \\
MAE      & .640          & .573          & .251          & .575          & .655          \\
-Two     & .651          & .637          & --            & .574          & .645          \\ \hline
-CW      & .645          & .631          & --            & .276          & .632          \\
+SW      & .643          & .629          & --            & .319          & .644          \\
-SM      & .654          & .639          & --            & \textbf{.584} & \textbf{.657} \\ \hline
B10m     & .656          & \textbf{.642} & --            & .577          & .655          \\
B100h    & .639          & .624          & --            & .552          & .634          \\
B960h    & \textbf{.658} & \textbf{.642} & --            & .567          & .647          \\ \hline
-Country & .655          & .639          & --            & .570          & --            \\
L        & .635          & .622          & --            & .554          & .603          \\
L960h    & .613          & .617          & --            & .540          & .539          \\ \hline
\end{tabular}
\label{tab:results_val}
\end{table}

\begin{table}[ht]
\centering
\caption{Results on the test split of the A-VB challenge dataset.}
\begin{tabular}{lccc}
\hline
         & High             & Culture       & Type          \\
Method   & CCC              & CCC           & UAR           \\ \hline
Baseline & .569             & .440          & .417          \\ \hline
-SM      & \textbf{.685}    & .515          & \textbf{.586} \\
B960h    & .679             & \textbf{.526} & .573          \\ \hline
\end{tabular}
\label{tab:results_test}
\end{table}

\section{Discussion}




Recapitulating the conducted experiments and considering the results, the following insights can be drawn:

\emph{Intermediate Tasks.} Determining the country of origin, optionally also the type of the vocal burst, before assessing the conveyed emotions like a rater of that country would, may be beneficial to a human rater, so that the emotional biases can be adjusted, but it is disadvantageous for neural networks -- the extracted features already encompass these biases and need not be handcrafted into.

\emph{Task Loss.} Varying the task losses did not improve the bad performance on the task \emph{Two}. Since the baseline shows double the performance here, the self-supervised learnt audio representations are unsuited for estimating valence and arousal. Therefore, the model learns to predict a rather constant output for \emph{Two}, so that the uncertainty in this task is reduced, minimising the penalising uncertainty term, but also maximising the task weight in the computation of the MTL loss. This can be seen in the learnt task weights, where \emph{Two} has 18-times (\emph{MSE}), respectively 4-times (\emph{MAE}), higher weight than the least weighted task (\emph{Country}), overshadowing other better-performing tasks. Removing \emph{Two} allows for higher variance and thereby better predictions in other tasks.

\emph{Weighting.} These experiments investigated different weighting techniques to counter imbalance in the training data. Removing the inversely proportional class weights in the calculation of the cross-entropy loss greatly reduces the performance in \emph{Type}. Extending the cross-entropy weights to the whole sample, to inversely proportional intra-batch weights depending on \emph{Country}, alleviates this only slightly. Inversely proportional class weights both in \emph{Type} and \emph{Country} to counter class imbalance, combined with the removal of sigmoid activation, thereby increasing the net's sensitivity to samples close to the boundaries of the value ranges, is the best approach.
Interestingly, mostly \emph{Type} is affected by these changes, which can be seen in its learnt task weight -- the uncertainty almost doubled.

\emph{Fine-Tuning Splits.} This addresses the assumption, that vocal bursts are not a word-based language, thereby we expect a decline in performance the more the pretrained net is fine-tuned using CTC on word-based labels. This decline is only slight for 10 minutes ($0.02\%$), but more so for 100 hours ($10.4\%$) of data used for fine-tuning. Interestingly, this does not hold for using all the data (960 hours) in fine-tuning. The predictions do differ, but on average, these differences cancel out. 

\emph{Network Size.} This investigates if the findings carry over if the net's feature extractor dimension is increased, or the output dimension is reduced, i.e. removing a task. Surprisingly, enlarging the feature extractor's dimension from 768 to 1024 decreases the performance noticeably. The more so, if it is fine-tuned on the whole LibriSpeech dataset after pretraining on Libri-light, where 960 hours resemble $1.6\%$ of data. Since both Libri-datasets are English-only, the large version of data2vec may overfit to the English language, hence reducing performance on non-English vocal bursts.
Removing the non-challenge task \emph{Country} decreases the performance of the remaining tasks. Thereby, using ``irrelevant'' data in multi-task learning does help.

\emph{Test results.} Regarding the test results in Table~\ref{tab:results_test}, there is still no clear difference visible between the pretrained and the fine-tuned base version of data2vec. In the tasks \emph{High} and \emph{Type}, the generalisability of the models hold, achieving similar results on the test set as on the validation set.
Note, that on task \emph{Culture}, both models decreased by 12 percentage points compared to validation results. Without the test labels it is not clear why, since the task \emph{High} performs well while sharing with task \emph{Culture} both the emotion-classification type and uncertainty ($+2.3\%$). All tasks' uncertainties are in the same order of magnitude, with task \emph{Culture} as the most uncertain and task \emph{Type} as the most certain ($-40\%$).
Hence, overfitting to \emph{Culture} seems unlikely, and it may be a sign of English-only pretraining and fine-tuning, which averages out for task \emph{High}.

\section{Conclusion \& Outlook}



We showed that no handcrafted features or assumptions have to be applied to achieve good results in multiple tasks simultaneously. Per-task ensembling and large structures are not needed, a single lightweight net for all tasks simultaneously suffices. Although being pretrained self-supervised only on English speech, data2vec is able to assess vocal bursts originating out of different (non-English) countries after fine-tuning it for fiveish hours.

The results may be further improved by enlarging the pretraining dataset by gathering multi-language samples in the wild and modifying the multi-task loss by penalising learning a constant prediction. Since the net's pretraining can be fine-tuned using CTC, it may be worthwhile to investigate datasets featuring other labels than words, that are more suited for vocal bursts, e.g. letters or phonemes.

\section*{Acknowledgements}
This work was partially funded by the KodiLL project (FBM2020, Stiftung Innovation in der Hochschullehre).

\bibliography{main}

\end{document}